\documentclass{article}
\usepackage{graphicx}
\usepackage{amsmath}
\begin{document}
\thispagestyle{empty}

\title{Scattering of solitons on resonance. Asymptotics and numeric simulations 
\thanks{This work was supported by grants RFBR 03-01-00716,
Leading Scientific Schools 1446.2003.1 and INTAS 03-51-4286.}}

\author{Oleg Kiselev\thanks{Institute of Math. USC
RAS; ok@ufanet.ru}, 
Sergei Glebov\thanks{Ufa State Petroleum Technical University; 
glebskie@rusoil.net}}

\date{}

\maketitle

\begin{abstract}
We investigate a propagation of solitons for
nonlinear Schr\"odinger equation under small driving force. The
driving force passes through the resonance. The process of scattering  on
the resonance leads to changing of number of solitons. After the
resonance the number of solitons  depends on the amplitude of the
driving force. The analytical results were obtained by WKB and matching method. We bring two examples of numeric simulations for verifying obtained analytical formulas. 
\end{abstract}

\font\Sets=msbm10
\def\Real{\hbox{\Sets R}}
\def\Complex{\hbox{\Sets C}}
\def\bb{\begin{equation}}
\def\ee{\end{equation}}
\def\pt{\partial}
\def\mod{\hbox{mod}}
\def\const{\hbox{const}}
\def\sgn{\hbox{sgn}}
\def\Arg{\hbox{Arg}}
\def\a{\alpha}
\def\b{\beta}
\def\d{\delta}
\def\G{\Gamma}
\def\g{\gamma}
\def\e{\epsilon}
\def\ve{\varepsilon}
\def\k{\kappa}
\def\l{\lambda}
\def\O{\Omega}
\def\o{\omega}
\def\th{t}
\def\s{\sigma}
\def\t{\tau}
\def\z{\zeta}

\noindent{\bf Introduction}

Nonlinear Schr\"odinger equation (NLSE) is a mathematical model for
wide class of wave phenomenons from the signal propagation into optical
fibre \cite{Kelley,Talanov} to the surface wave propagation
\cite{Zakharov}. This equation is integrable by inverse scattering
transform method \cite{TeoriyaSolitonov} and can be considered as an
ideal model equation. The perturbations of this ideal model lead to
nonintegrable equations. Here we consider such nonintegrable example
which is NLSE  perturbed by the small driving force.
\bb
i \pt_t \Psi + \pt_{x}^2 \Psi + |\Psi|^2 \Psi = \ve^2 f
e^{iS/\ve^2},\qquad 0< \ve \ll 1. \label{sh}
\ee
\par
The perturbed NLSE in form (\ref{sh}) is relative to describing of nonlinear effects of optical soliton propagation in the presence of an input fast oscillating forcing beam \cite{1}.
\par
The most known class of the solutions of NLSE is solitons
\cite{TeoriyaSolitonov}. The structure of this kind of solutions is
not changed in a case of nonperturbed NLSE. The perturbations
usually  lead to  modulation of parameters of solitons \cite{Kaup,Karpman-Maslov}, for driven NLSE see also \cite{Bar}. 
Number of solitons does not change.
\par
In this work we investigate a new effect called the scattering of
solitons on the local resonance. Typical picture for this process may be obtained by numerical simulations (see fig.1).
\begin{figure}[ht]{\includegraphics[width=16cm,height=12.97cm]{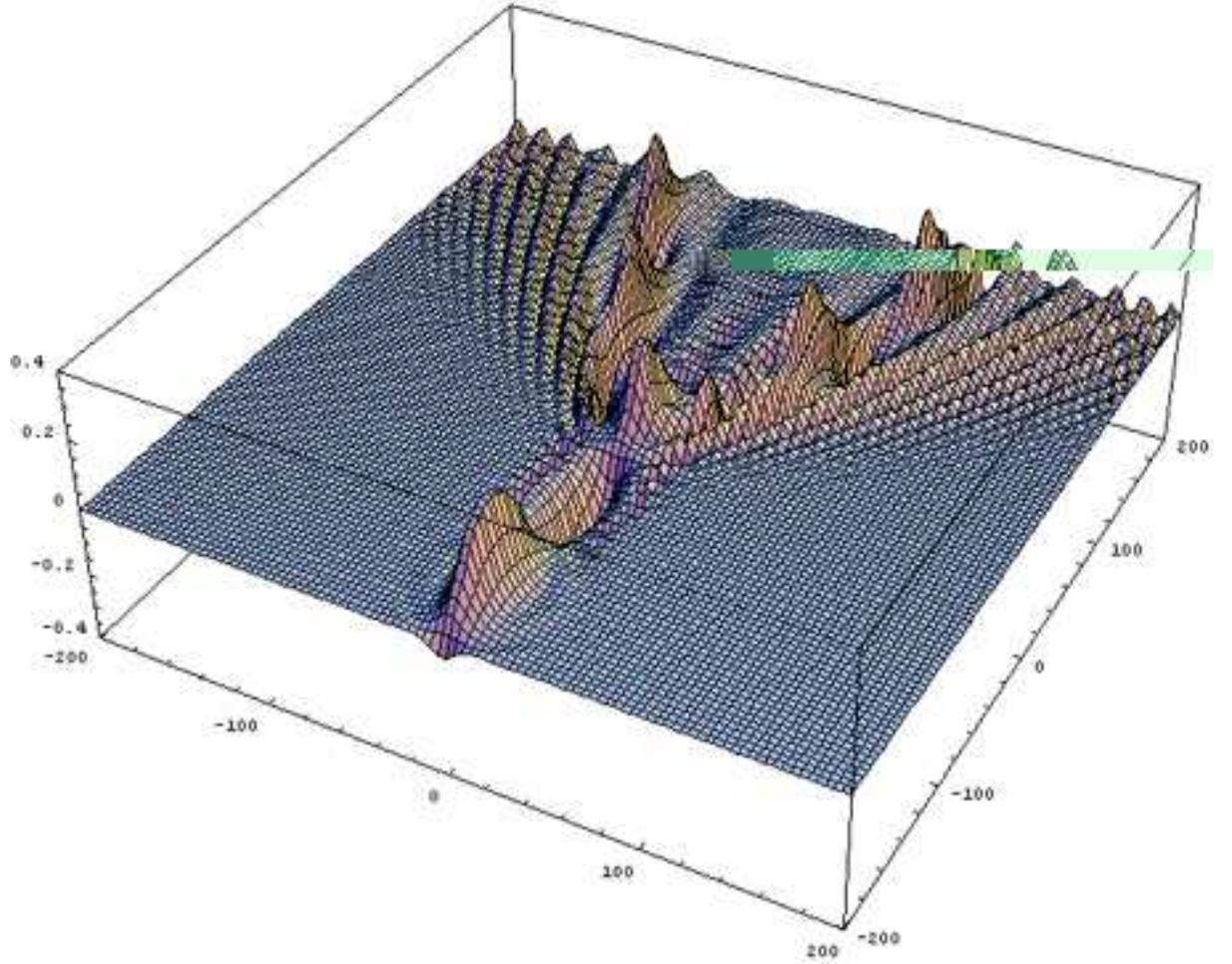}}
\caption{This picture shows the scattering process of one soliton to two solitons for equation (\ref{sh}), where  amplitude of external force $\ve=0.1$, $S=0.005t^2$, $f= 2\sqrt{2}\cosh^{-1}(0.2x)+ $  $2\sqrt{2}\exp(0.2ix)\cosh^{-1}(0.2x-5)+$  $2\sqrt{2}\exp(-0.2ix)\cosh^{-1}(0.2x+5),$ initial data is pure soliton of NLSE at $\Psi|_{t=-200}=0.2\sqrt{2}\cosh^{-1}(0.2x)$ , resonance curve is $t=0$.}
\end{figure}
\par
Let us explain the terms 'scattering' and 'resonance' which is used in the work. Usually one says 'scattering' for process when a wave is changed after obstacle. The same process is studied in our work for solitons which are nonlinear waves of NLSE. The obstacle is a line where the external force has a resonance in the equation for first-order term in the asymptotic solution for perturbed NLSE. Under 'resonance' we understand standard phenomenon of grows solution because of oscillating external force. The resonance phenomenon takes place in a thin domain near  some curve usually called a resonance curve. So we use the term 'local resonance'.
\par
We consider the process of scattering in
detail and obtain the connection formula between pre-resonance and
post-resonance solutions. In general case the passage through
the resonance leads to changing of the number of solitons. This effect
is based on the phenomenon of soliton generation due to the passage through the resonance
by the external driving force \cite{Glebov-Kiselev-Lazarev}.
\par
We found that the scattering of solitary waves on the resonance 
is a general effect for the wave propagation in a nonlinear media. 
In this work we investigate this effect for the simplest model. 
It allows to show the essence of this effect without unnecessary details.
\par
This paper has the following structure. The first section contains
the statement of the problem and the main result. The sections 2-4 are reviewed asymptotic analysis of the problem which was developed in \cite{SGOK}. The second section
contains the asymptotic construction in the pre-resonance domain. In
the third section we construct  the asymptotic solution in the
neighborhood of the resonance curve. The fourth section of the paper
is devoted to construction of the post-resonance asymptotics.
Asymptotics are constructed by multiple scale method \cite{JK} and
matched \cite{Il'in}. At last five section contains the results of numerical simulation which justify the obtained asymptotic formulas. 
\par

\section{Statement of the problem and  result}

We study the perturbed NLS equation (\ref{sh}) with a special phase of the driving force: $S/\ve^2 = \ve^2 t^2$. The amplitude $f=f(\ve x)$ is a smooth and rapidly vanished function. The small parameter in the right hand side of the equation is defined as $\ve^2$ only for a convenience. 
\par
The goal of our work is the study of a slow evolution of the solution with a small amplitude for the equation (\ref{sh}). In the general case the small amplitude solutions are defined by the linear Schr\"odinger equation. However there exists a magic relation between scales of independent variables of a carrier wave and an amplitude of its envelope function which leads to the nonlinear Schr\"odinger equation for the envelope function of the leading-order term of the asymptotics. This relation was observed for different physical problems in pioneer works \cite{Kelley,Talanov,Zakharov}. An example of such relation give the following substitution
$$
\Psi=\ve U(t_2,x_1,\ve),\quad t_2=\ve^2 t,\ \  x_1=\ve x,
$$
then the equation (\ref{sh}) has the form:
\bb
i \pt_{t_2} U + \pt_{x_1}^2 U + |U|^2 U = \ve^{-1} f
e^{iS/\ve^2}.\label{sh2}
\ee
The strong perturbation with rapid phase for NLSE may be considered as a model for the high-frequency heating of plasma \cite{Petv-Pokh} and leads to the phenomenon of the scattering of solitons.  
\par
It is known if $f\equiv0$ then there exists a soliton solution of the  NLSE. We'll show that for $f\not\equiv0$ the number of solitons may be changed due to passage through the resonant curve. The resonant curve defines as a line where the frequency of the driving force is equal to the eigenfrequency of linearized Schr\'odinger equation. In our case this curve is $t=0$. 
After passage the resonance the number of solitons  depends on amplitude of the perturbation on the resonance curve $t=0$. 
\par
In the simplest case the phase is linear function with respect to $t$
$S/\ve^2=\omega t,\ \omega=\const$. In general situation the
constant frequency of the driving force does not lead to the scattering
of solitons. Let us investigate the driving force with slowly
varying frequency. The most simplest dependence on $t$ for $\omega$
has a form $\omega=\ve^2 t/2$. Hence $S=\ve^4t^2/2$. Namely the equation with this form of the phase function of the driving force is studied in this work. The amplitude $f$ of the driving
force admit an additional dependency on $\ve^2 t$ but it leads to
complicated formulas and no more.

\par
Let us formulate the result of this  work. Below we use the
following variables $x_j=\ve^j x,\ t_j=\ve^j t,\ j=1,2$.
\par
Let the asymptotic solution of (\ref{sh}) be
$$
\Psi(x,t,\ve)=\ve \stackrel{1}{u}(x_1,t_2)+O(\ve^2)\quad
\mbox{as}\quad -c< t_2 < 0,
$$
where $c=\const>0$ and  $\stackrel{1}{u}(x_1,t_2)$ satisfies
$$
\pt_{t_2}\stackrel{1}{u}+\pt_{x_1}^2 \stackrel{1}{u}+|
\stackrel{1}{u}|^2 \stackrel{1}{u}=0
$$
and initial condition
$$
\stackrel{1}{u}|_{t_2=t^0}=h_1(x_1),\quad t^0=\const < 0.
$$
Then in the domain $0<t_2<c$ the asymptotic solution of (\ref{sh}) has
a form
\begin{eqnarray}
\Psi(x,t,\ve)=\ve \stackrel{1}{v}(x_1,t_2)+O(\ve^2),
\end{eqnarray}
where $\stackrel{1}{v}(x_1,t_2)$ is a solution of NLSE with initial
condition
\bb
\stackrel{1}{v}|_{t_2=0}=\stackrel{1}{u}(x_1,0)+(1-i)\sqrt{\pi}f(x_1).
\label{initialCondition}
\ee
\par
Formula (\ref{initialCondition}) is the main result of the paper. This formula connects the main order term $\stackrel{1}{u}$ of the asymptotic solution before the scattering, external driving force $f$ and initial  condition $\stackrel{1}{v}$ for the main order term of the asymptotic solution after the scattering. This formula is derived in the end of section 4. 
\par
Let us explain the result for soliton solution. If in the domain
$-c<t_2<0$ the solution has   $N$-soliton form then in the domain
$0<t_2<c$ the number of solitons is defined by initial condition (\ref{initialCondition}).
\par
Given analysis is valid at $|t_2|\le C$, for $\forall C=\const$. When $|\ve^2t|\gg1$ the perturbation force will modulate the parameters of the soliton solution, see \cite{Kaup,Karpman-Maslov}.

\section{Incident wave}
\par
In this section we construct the asymptotic solution of equation
(\ref{sh}) in pre-resonance domain. This solution contains two
parts. The first part is a specific solution of the nonhomogeneous
equation. This solution oscillates with the frequency of the driving
force. The amplitudes are determined by an algebraic equations. The
second part of the solution is a solution of the homogeneous
equation. The solution contains an undefined function due to
integration. This undefined function usually determines by initial
condition for the Cauchy problem.
\par
We construct the formal asymptotic solution in the WKB-like form
\begin{eqnarray}
\Psi(x,t,\ve)=\ve \stackrel{1}{u}(x_1,t_2)+\ve^3\stackrel{3}{u}(x_1,t_2)+
\ve^{2}\stackrel{2}{B}(x_1,t_2)\exp(iS/\ve^2)+ \nonumber\\
\ve^4\bigg(\stackrel{4}{B}_{1}(x_1,t_2)\exp(iS/\ve^2)+
\stackrel{4}{B}_{-1}(x_1,t_2)\exp(-iS/\ve^2)\bigg)
\label{external-anzats-2}\\
+\ve^{5}\stackrel{5}{B}_2(x_1,t_2)\exp(2iS/\ve^2).\nonumber
\end{eqnarray}
\par
To determine the coefficients of the asymptotics substitute
(\ref{external-anzats-2}) into equation (\ref{sh}). It yields
\begin{eqnarray*}
\ve^2\bigg(-S'\stackrel{2}{B}-f\bigg)\exp(iS/\ve^2)+
\ve^3\bigg(i\stackrel{1}{u}_{t_2}+\stackrel{1}{u}_{x_1x_1}+
|\stackrel{1}{u}|^2\stackrel{1}{u}\bigg)
\\
+\ve^{4}\bigg(\bigg(-S'\stackrel{4}{B}_{1}+
i\stackrel{2}{B}_{t_2}+
\stackrel{2}{B}_{x_1x_1}+
2|\stackrel{1}{u}|{}^2\stackrel{2}{B}\bigg)\exp(iS/\ve^2)+\\
+\bigg(S'\stackrel{4}{B}_{-1}+
\stackrel{1}{u}{}^2\stackrel{2}{B}{}^*\bigg)\exp(-iS/\ve^2)\bigg)
\\
\ve^5\bigg(
i\stackrel{3}{u}_{t_2}+\stackrel{3}{u}_{x_1x_1}
+2|\stackrel{1}{u}|^2\stackrel{3}{u}+
\stackrel{1}{u}{}^2\stackrel{3}{u}{}^*+
2\stackrel{1}{u}|\stackrel{2}{B}|{}^2+\nonumber\\
\bigg(-2S'\stackrel{5}{B}_{2}+
\stackrel{1}{u}{}^*\stackrel{2}{B}{}^2\bigg)\exp(2iS/\ve^2)\bigg)
= \ve^6R(t_2,x_1).
\end{eqnarray*}
\par
The residue part of the asymptotics  has a form
\bb
R(t_2,x_1)=O(|\stackrel{2}{B}|^3+\ve^3|\stackrel{3}{u}|^3+\ve^6|\stackrel{4}{B}_{1}|^3+\ve^6|\stackrel{4}{B}_{-1}|^3+\ve^9|\stackrel{5}{B}_2|^3).
\label{residuePart1}
\ee
\par
Collect the terms with the same order of $\ve$ up to  the order of
$\ve^5$ and reduce similar  terms. It yields  differential equations
for $\stackrel{1}{u}, \stackrel{3}{u}$ and algebraic equations for
$\stackrel{2}{B}$, $\stackrel{4}{B}_{\pm1}$ and
$\stackrel{5}{B}_{2}$.
\begin{eqnarray}
i\stackrel{1}{u}_{t_2}+\stackrel{1}{u}_{x_1x_1}+
|\stackrel{1}{u}|^2\stackrel{1}{u}=0,\label{u1}\\
i\stackrel{3}{u}_{t_2}+\stackrel{3}{u}_{x_1x_1}+
2|\stackrel{1}{u}|^2\stackrel{3}{u}+
\stackrel{1}{u}{}^2\stackrel{3}{u}{}^*=
-2|\stackrel{2}{B}|^2\stackrel{1}{u},\label{u3}
\end{eqnarray}
\begin{eqnarray}
-S'\stackrel{2}{B}=f,\label{b2} \\
-S'\stackrel{4}{B}_1=i\stackrel{2}{B}_{t_2}+\stackrel{2}{B}_{x_1x_1}+
|\stackrel{1}{u}|^2\stackrel{2}{B}, \label{b4}\\
S'\stackrel{4}{B}_{-1}=-\stackrel{1}{u}{}^2 \stackrel{2}{B}{}^*,\label{b41}\\
-2S'\stackrel{5}{B}_2=\stackrel{1}{u}{}^*\stackrel{2}{B}{}^2.\label{b5}
\end{eqnarray}
\par
The functions $\stackrel{1}{u},  \stackrel{3}{u}$  are uniquely
determined by initial conditions at the moment $t_2=t^0$. We suppose
that $t^0=\const<0$ and
$$
\stackrel{1}{u}|_{t_2=t^0} = h_1(x_1);\quad
\stackrel{3}{u}|_{t_2=t^0} = h_3(x_1);
$$
where functions $h_1, h_3$ are smooth and rapidly vanish as $|x_1|\to
\pm\infty$. 
\par 
It's known the solutions $\stackrel{1}{u}$ and $\stackrel{3}{u}$ exist for bounded values of  $t_2$, see \cite{DEGM,Keen-McL}. 
\par{\bf Remark} The solution of the equation for $\stackrel{3}{u}$ contains growing terms as $t_2\to\infty$, see for example \cite{Keen-McL}. These terms are secular as $t_2\sim\ve^{-1}$.  But we do not consider such long times in this work.
\par
The coefficients of the representation (\ref{external-anzats-2})
have a singularity as $S'\to 0$. The order of singularity of
$\stackrel{j}{B}_{k}$ is easy calculated.
$$
\stackrel{2}{B}=O(t^{-1}),\quad \stackrel{4}{B}_1=O(t^{-3}).
$$
\par
To determine the asymptotics of $\stackrel{3}{u}$ as $t_2\to-0$
we construct the solution of the form
\bb
\stackrel{3}{u}=t_2^{-1}\stackrel{3}{u}{}^{(-1,0)}(x_1,t_2)+
\ln|t_2|\stackrel{3}{u}{}^{(0,1)}(x_1,t_2)+
t_2\ln|t_2|\stackrel{3}{u}{}^{(1,1)}(x_1,t_2)+
\widehat{\stackrel{3}{u}}(x_1,t_2).
\label{asu3}
\ee
Substitute this representation into equation (\ref{u3})
and collect the terms of the same order with respect to $t_2$. It
yields equations for coefficients of the asymptotics (\ref{asu3})
\begin{eqnarray}
\stackrel{3}{u}{}^{(-1,0)}=i2|f|^2\stackrel{1}{u},\nonumber\\
\stackrel{3}{u}{}^{(0,1)}=-iL(\stackrel{3}{u}{}^{(-1,0)}),\nonumber\\
\stackrel{3}{u}{}^{(1,1)}=-iL(\stackrel{3}{u}{}^{(0,1)}),\nonumber\\
L(\widehat{\stackrel{3}{u}})=it_2\ln|t_2|L(\stackrel{3}{u}{}^{(1,1)})+
i\stackrel{3}{u}{}^{(1,1)}.\label{hat-u3-eq}
\end{eqnarray}
Here  $L(u)$ is a linear operator of the form
$$
L(u)=i\pt_{t_2}u+\pt_{x_1}^2 u+2|\stackrel{1}{u}|^2 u +\stackrel{1}{u}{}^2u^*.
$$
Functions $\stackrel{3}{u}{}^{(-1,0)}$,
$\stackrel{3}{u}{}^{(0,1)}$ and $\stackrel{3}{u}{}^{(1,1)}$ are
determined from algebraic equations. These functions are bounded
as $-\const<t_2\le0,\  \const>0$.
\par
The function $\widehat{\stackrel{3}{u}}$ is a solution of
nonhomogeneous  linearized Schrodinger equation. The right hand side
of the equation is a smooth function as $-\const<t_2\le0,\
\const>0$. The solution of this equation can be obtained using
results of \cite{Keen-McL}. In particularly if $\stackrel{1}{u}$ is
N-solitons solution of NLSE then exists the bounded solution of
nonhomogeneous  linearized Schrodinger equation (\ref{hat-u3-eq}) as
$-\const<t_2\le0,\  \const>0$.
\par
The asymptotic form (\ref{external-anzats-2}) allows to solve equation (\ref{sh}) up to the order $\ve^6$. To obtain more accurate approximation one have  to include  terms without fast oscillating of the order $o(\ve^3)$ into asymptotic solution (\ref{external-anzats-2}). Therefore we define the domain of validity of (\ref{external-anzats-2}) by following relation:
$$
\ve^6R(t_2,x_1)=o(\ve^3),\quad \ve\to0.
$$
Coefficients of (\ref{external-anzats-2}) have  singularity at
$t_2=0$.  The residue part increases as
$t_2 \to 0$. From formulas (\ref{residuePart1}) and (\ref{u1})--(\ref{b5}) one can easily obtain the behaviour of the residue part: 
$$
R(t_2,x_1)=O(t_2^{-3}+\ve^6 t_2^{-9}),\quad t_2\to-0.
$$
It yields the domain of validity of (\ref{external-anzats-2})
$$
-t_2\gg\ve \quad \hbox{or} \quad -t \gg \ve^{-1}.
$$

\section{Scattering}

In the neighborhood of the point $t_2=0$ the frequency of the
driving force becomes resonant.  Formally it means representation
(\ref{external-anzats-2}) is not valid.
\par
In this part of the work we construct another representation for the
solution of equation (\ref{sh}). This representation is valid in the
neighborhood of the resonance line $t_2=0$.
\begin{eqnarray}
\Psi(x,t,\ve)=\ve\stackrel{1}{w}(x_1,t_1)\big)+
\ve^{2}\stackrel{2}{w}(x_1,t_1)+\nonumber\\
\ve^3\ln\ve\stackrel{3,1}{w}(x_1,t_1)+
\ve^{3}\stackrel{3}{w}(x_1,t_1) \qquad
 \ve \to 0. \label{internal-anzats}
\end{eqnarray}
Here we use a new scaled variable $t_1=t_2/\ve$. Representation
(\ref{internal-anzats}) is matched with (\ref{external-anzats-2}).
It means these formulas are equivalent up to value $o(\ve^5)$ as
$t_2\to-0$. The coefficients of (\ref{internal-anzats}) are determined 
by ordinary differential equations (\ref{inw0}), (\ref{inw1}), (\ref{inw2}) 
and matching conditions.
\par
To obtain the behaviour of the coefficients of
(\ref{internal-anzats}) as $t_1\to-\infty$ match
(\ref{internal-anzats}) with (\ref{external-anzats-2}). Write
(\ref{external-anzats-2}) in terms of $t_1$
$$
\Psi(x,t,\ve)=\ve\bigg(\stackrel{1}{u}(x_1,0)-\big(t_1^{-1}f+
it_1^{-3}f\big)\exp(iS/\ve^2)\bigg)
+
$$
$$
\ve^2\bigg(\pt_{t_2}\stackrel{1}{u}(x_1,t_2)|_{t_2=0} t_1+
t_1^{-1}i|f|^2\stackrel{1}{u}(x_1,0)+O\big(t_1^{-2}\big)\bigg)+
$$
$$
\ve^3\ln\ve\bigg(-iL(2i|f|^2\stackrel{1}{u})|_{t_2=0}+o(1)\bigg)+
$$
$$
 +\ve^3\bigg({1\over2}\pt_{t_2}^2\stackrel{1}{u}(x_1,t_2)|_{t_2=0}
t_1^2+\widehat{\stackrel{3}{u}}(x_1,0)+o(1)\bigg), \quad
1\ll-t_1\ll\ve^{-1},\,\,\ve\to0.
$$
\par
To obtain equations for coefficients of (\ref{internal-anzats})
substitute (\ref{internal-anzats}) into equation (\ref{sh}). It
yields
\begin{eqnarray*}
\ve^2\bigg((\pt_{t_1}\stackrel{1}{w}-f\exp(iS/\ve^2)\bigg)+
\ve^3\bigg(\pt_{t_1}\stackrel{2}{w}+ \pt_{x_1}^2\stackrel{1}{w}+
\gamma|\stackrel{1}{w}|\stackrel{1}{w}\bigg) +\\
\ve^4\bigg(\pt_{t_1}\stackrel{3}{w}+ \pt_{x_1}^2\stackrel{2}{w}+
+\stackrel{1}{w}{}^2\stackrel{2}{w}{}^* +
2\gamma|\stackrel{1}{w}|\stackrel{2}{w}\bigg)=\ve^5\rho(t_1,x_1,\ve).
\end{eqnarray*}
The function $\rho(t_1,x_1,\ve)$ can be represented in the form
$$
\rho(t_1,x_1,\ve)=O(|\stackrel{1}{w}|^2\stackrel{3}{w}+\pt_{x_1}^2\stackrel{3}{w}
+\ve|\stackrel{2}{w}|^3+\ve^4|\stackrel{3}{w}|^3).
$$
Collect the terms of the same order with respect to $\ve$. As
result we obtain the equations for coefficients of (\ref{internal-anzats}).
\begin{eqnarray}
i\pt_{t_1}\stackrel{1}{w}=f\exp(it_1^2/2). \label{inw0}
\end{eqnarray}
The matching conditions give
$\stackrel{1}{w}=\stackrel{1}{u}(x_1,0)\,\,\, t_1\to-\infty$. The
solution of this problem is represented in terms of Fresnel
integral
\bb
\stackrel{1}{w}=\stackrel{1}{u}(x_1,0) -if(x_1)\int_{-\infty}^{t_1}\exp(i\theta^2/2)d\theta.
\label{leading-order-of internal-exp}
\ee
Equations for higher-order terms are
\bb
i\pt_{t_1}\stackrel{2}{w}=-\pt_{x_1}^2\stackrel{1}{w}+
|\stackrel{1}{w}|^2\stackrel{1}{w}, \label{inw1}
\ee
\bb
i\pt_{t_1}\stackrel{3,1}{w}=0, \label{inw21}
\ee
\bb
i\pt_{t_1}\stackrel{3}{w}=-\pt_{x_1}^2\stackrel{2}{w}-
2|\stackrel{1}{w}|^2\stackrel{2}{w}-
\stackrel{1}{w}{}^2\stackrel{2}{w}{}^*. \label{inw2}
\ee
\par
The higher-order terms satisfy fist order ordinary differential
equations with respect to $t_1$. The spatial variable $x_1$ is a
parameter in these equations. The solutions of these equation are
uniquely defined by terms of the order of $1$ in asymptotics as
$t_1\to-\infty$. The asymptotics as $t_1\to-\infty$ is obtained
by matching
$$
\stackrel{2}{w}=\pt_{t_2}\stackrel{1}{u}(x_1,t_2)|_{t_2=0} t_1+o(1);
$$
$$
\stackrel{3,1}{w}=-iL(2i|f|^2\stackrel{1}{u})|_{t_2=0}+o(1),
$$
$$
\stackrel{3}{w}={1\over2}\pt_{t_2}^2\stackrel{1}{u}(x_1,t_2)|_{t_2=0} t_1^2+\widehat{\stackrel{3}{u}}(x_1,0)+o(1).
$$
To determine the behaviour of the solution after resonance we need
to calculate the asymptotics as $t_1 \to +\infty$ of the
coefficients for representation (\ref{internal-anzats}).
Calculations give
$$
\stackrel{1}{w}(x_1,t_1)=\stackrel{1}{u}(x_1,0)-i f(x_1)\bigg[ic_1 +
{\exp(it_1^2/2)\over it_1}+O(t_1^{-3})\bigg],
$$
where $c_1 = (1-i)\sqrt{\pi}$.
\par
Denote by
$$
\stackrel{1}{w}(x_1,t_1)|_{t_1\to\infty}=\stackrel{1}{w}_0(x_1).
$$
The function $\stackrel{2}{w}(x_1,t_1)$ has the asymptotics of the
form
$$
\stackrel{2}{w}(x_1,t_1)=t_1\stackrel{2}{w}{}_{1}(x_1) + \stackrel{2}{w}{}_0(x_1)+ g_1(x_1){\exp(it_1^2/2)\over
it_1^2} +O(t_1^{-4}),
$$
where
$$
\stackrel{2}{w}{}_1=-\pt_{x_1}^2\stackrel{1}{w}_0+|\stackrel{1}{w}_0|^2\stackrel{1}{w};
$$
$$
\stackrel{2}{w}{}_{0}(x_1)=\lim_{t_1\to\infty}\bigg(
\int_{-\infty}^{t_1} \big[\pt_{x_1}^2\stackrel{0}{w}(x_1,\theta)+
|\stackrel{0}{w}(x_1,\theta)|^2\stackrel{0}{w}(x_1,\theta)\big]d\theta
-\stackrel{2}{w}_{1}t_1\bigg),
$$
$g_1(x_1)=k_1 \pt_{x_1}f+k_2|f|^2 f$, $k_1$ and $k_2$ are
constants.
$$
\stackrel{3}{w}(x_1,t_1)=t_1^2\stackrel{3}{w}{}_{2}(x_1)
+ o(t_1^2),\quad t_1\to\infty,
$$
where
$$
\stackrel{3}{w}{}_2(x_1)=i\left(\pt_{x_1}^2\stackrel{2}{w}_1
+2|\stackrel{1}{w}_0|^2\stackrel{2}{w}_1 +
\stackrel{1}{w}{}_0^2\stackrel{2}{w}{}_1^*\right),
$$
\par
The domain of validity for (\ref{internal-anzats}) is defined by the same way as was shown  in previous section. We require the following relation is valid
$$
\ve^5\rho(t_1,x_1,\ve)=o(\ve^2),\quad t_1\to\infty.
$$
The determined above behaviour of coefficients of asymptotics
(\ref{internal-anzats}) give the domain of validity for (\ref{internal-anzats}) 
$$
|t_1| \ll \ve^{-1} \quad \hbox{or} \quad |t| \ll \ve^{-2}.
$$

\section{Scattered wave}

In this section we construct the asymptotic solution of equation
(\ref{sh}) after passage through the resonance.  The leading-order term of the
solution satisfies NLSE and depends on $x_1,t_2$ as well as before
resonance. But this leading-order term  is determined by another
solution of NLSE which contains generally speaking another number of
solitons. 
\par
We construct the asymptotic solution of the form
\begin{eqnarray}
\Psi(x,t,\ve)=\ve\stackrel{1}{v}(x_1,t_2)+
\ve^2\stackrel{2}{v}(x_1,t_2)+
\nonumber\\
\ve^2\stackrel{2}{A}(t_2,x_1)\exp(iS/\ve^2)+\ve^4(\stackrel{4}{A}_1(t_2,x_1)\exp(iS/\ve^2)+
\nonumber\\
\stackrel{4}{A}_{-1}(t_2,x_1)\exp(-iS/\ve^2)).
\label{external-anzats2}
\end{eqnarray}
\par
Substitute this representation into  (\ref{sh}):
\begin{eqnarray*}
\ve^2(-S'\stackrel{2}{A}-f)\exp(iS/\ve^2) +\ve^3(\pt_{t_2}\stackrel{1}{v}+\pt_{x_1}^2\stackrel{1}{v}+ |\stackrel{1}{v}|^2\stackrel{1}{v})+
\\
\ve^4(\pt_{t_2}\stackrel{2}{v}+\pt_{x_1}^2\stackrel{2}{v}+2 |\stackrel{1}{v}|^2\stackrel{2}{v}+ \stackrel{1}{v}^2\stackrel{2}{v}{}^*+
\\ (-S'\stackrel{4}{A}_{1}+\pt_{t_2}\stackrel{2}{A}+ \pt_{x_1}^2\stackrel{2}{A}+2|\stackrel{1}{v}|^2\stackrel{2}{A})\exp(iS/\ve^2)+\\ (S'\stackrel{4}{A}_{-1}+\stackrel{1}{v}{}^2\stackrel{2}{A}{}^*)\exp(-iS/\ve^2))=\ve^5 r(t_2,x_1,\ve).
\end{eqnarray*}
Here $r(t_2,x_1,\ve)$ depends on coefficients of the asymptotics
(\ref{external-anzats2}). This dependence is easy calculated. The
coefficients $\stackrel{2}{A}$, $\stackrel{4}{A}_1$ and 
$\stackrel{4}{A}_{-1}$ have singularity on the resonance curve. To
determine the domain of validity of (\ref{external-anzats2}) we need
to derive the explicit formula for $r$
$$
r(t_2,x_1,\ve)=O(1+\ve|\stackrel{2}{A}|^3+\ve\ln|\ve|+\ve^7(|\stackrel{4}{A}_1|^3
+|\stackrel{4}{A}_{-1}|^3)).
$$
\par
Collect the terms of the same order of small parameter and the same
exponents. It yields the equations for coefficients of representation
(\ref{external-anzats2}).
\bb
\pt_{t_2}\stackrel{1}{v}+\pt_{x_1}^2\stackrel{1}{v}+ |\stackrel{1}{v}|^2\stackrel{1}{v}=0;
\label{sh-2}
\ee
\bb
\pt_{t_2}\stackrel{2}{v}+\pt_{x_1}^2\stackrel{2}{v}+2 |\stackrel{1}{v}|^2\stackrel{2}{v}+ \stackrel{1}{v}{}^2\stackrel{2}{v}{}^*=0
\label{lsh-1}
\ee
$$
-S'\stackrel{2}{A}=f;
$$
$$
-S'\stackrel{4}{A}_{1}=-\pt_{t_2}\stackrel{2}{A}- \pt_{x_1}^2\stackrel{2}{A}-2|\stackrel{1}{v}|^2\stackrel{2}{A};
$$
$$
S'\stackrel{4}{A}_{-1}=-\stackrel{1}{v}{}^2\stackrel{2}{A}{}^*.
$$
\par
Initial conditions for differential equations for $\stackrel{1}{v}$
are obtained by matching. These conditions are evaluated on the
resonance curve $t_0=0$.
\bb
\stackrel{1}{v}|_{t_2=0}=\stackrel{1}{u}(x_1,0)+(1-i)\sqrt{\pi}f(x_1);
\label{ic1}
\ee
\bb
\stackrel{2}{v}|_{t_2=0}=\stackrel{2}{w}_{0}(x_1).
\label{ic2}
\ee
\par
The residue part $\ve^5r(t_2,x_1,\ve)=o(\ve^2)$ as $t_2\gg\ve$. This
condition is determined the domain of validity for
(\ref{external-anzats2}).
\par
Formula (\ref{ic1}) is connection formula for the leading-order term of the asymptotic 
solution before and after the resonance. Additional term $(1-i)\sqrt{\pi}f(x_1)$ leads
to changing of the solution after passage through the resonance.

\section{Numerical justification of asymptotic analysis}

In this section we justify our asymptotic formula (\ref{initialCondition}).  
Let us consider the pure soliton initial condition for equation (\ref{sh}):
\bb
\Psi(x,t,\ve)\vert_{t=t_0} ={{2\sqrt{2}\ve\eta\exp\{-i2c\ve x-4(c^2-\eta^2)t_0\ve^2 \}}\over{\cosh(2\eta\ve x + 8c\eta \ve^2 t_0+s)}}
\ee
According of our analytical results this initial condition leads to one soliton solution as the leading-order term of the asymptotic solution:
$$
\stackrel{1}{u}(x_1,t_2)={{2\sqrt{2}\eta\exp\{-i2c x_1 -4i(c^2-\eta^2)t_2 \}}\over{\cosh(2\eta  x_1 + 8c\eta t_2+s)}}.
$$
This soliton propagates up to the resonance curve $t=0$. 
\par
To annihilate this soliton on the resonance curve one may choose the specific 
form of the amplitude of the perturbation such that the left hand side of  relation (\ref{initialCondition}) equals zero:
$$
0=\stackrel{1}{u}(x_1,0)+(1-i)\sqrt{\pi}f(x_1).
$$ 
Hence 
$$
f(x_1)={-(1+i)\over2\sqrt{\pi}}\stackrel{1}{u}(x_1,0)=
{-(1+i)\over\sqrt{\pi}} {{\sqrt{2}\eta\exp\{-i2c x_1\}}\over{\cosh(2\eta  x_1 + s)}}
$$
To illustrate this  by numerical simulations we choose $\ve=0.1, \eta=1, s=0, c=0, t^0=-200$. Then the original equation (\ref{sh})  has the form
$$
i\pt_t\Psi + \pt_x^2 \Psi +|\Psi|^2\Psi = 0.01 f \exp\{i 0.005 t^2\}.
$$
Initial condition is
$$
\Psi\vert_{t=-200} ={0.2\sqrt{2}\over{\cosh(0.2x)}},
$$
and amplitude of the perturbation is
$$
f = {-(1+i)\over\sqrt{\pi}} {{\sqrt{2}}\over{\cosh(0.2 x)}}
$$
The numerical simulations of annihilation process for soliton of NLSE are presented on the following figure. This justifies the formulas obtained  above by matching method.  
\par
\begin{figure}[ht]{\includegraphics[width=16.02cm,height=12.98cm]{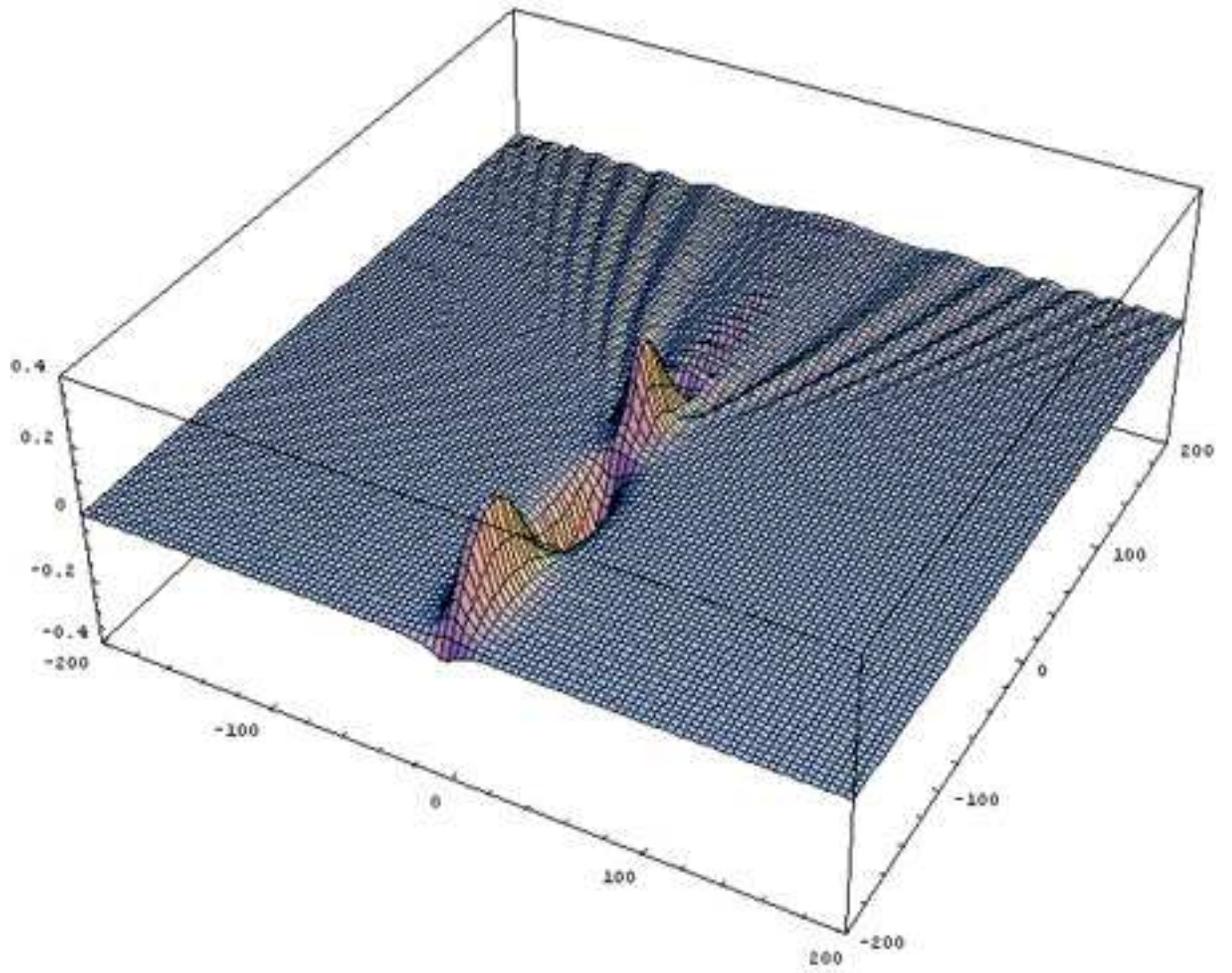}}
\caption{Annihilation of soliton.}
\end{figure}

\section{Summary}
\par
We have studied  the scattering process on the resonance and obtained the way for the control the parameters of scattered solitons. Using formula (\ref{initialCondition}) one can choose the form of the perturbation to obtain the scattered pattern of the given form. The process of the scattering on the local resonance is an universal phenomenon for waves propagate in dispersion media. We believe that our work can be useful for further understanding and description of solitary waves evolution under perturbation. 

\par
{\bf Acknowledgments.} We are grateful to  I.V. Barashenkov, L.A. Kalyakin and B.I.Suleimanov for helpful comments. This work was supported by RFBR 03-01-00716, grant for Sci. Schools 1446.2003.1 and INTAS 03-51-4286.


\begin{thebibliography}{c}
\bibitem{Kelley}
Kelley P.L. Self-focusing of optical beams. Phys.Rev.Lett., 1965,
v.15, 1005-1008.
\bibitem{Talanov}
Talanov V.I. O samofokusirovke malykh puchkov v nelineinykh sredakh,
Pis'ma v ZhETF, 1965, n2, 218-222.
\bibitem{Zakharov}
Zakharov V.E. Ustoichivost' periodicheskikh voln s konechnoi
amplitudoi na poverkhnosti glubokoi zhidkosti. Zhurnal prikladnoi
mekhaniki i tekhnicheskoi fiziki, 1968, n2, 86-94.
\bibitem{TeoriyaSolitonov}
Zakharov V.E., Manakov S.V., Novikov S.P., Pitaevskii L.P. Teoriya
solitonov: metod obratnoi zadachi. M.:Nauka, 1980.
\bibitem{1} Lugiato L.A. and Lefever R., Spatial Dissipative Structures in Passive Optical Systems. Phys. Rev. Lett. 58, 2209 (1987).
\bibitem{Kaup}
Kaup D.J. A perturbation expansion for  the Zakharov-Shabat inverse
scattering transform. SIAM J.on Appl.Math., 1976, v. 31, 
121--133.
\bibitem{Karpman-Maslov}
Karpman V.I., Maslov E.I.   Teoriia vozmusheniy dlia solitonov
ZhETPh, 1977, t.73, 537--559.
\bibitem{Bar}
I.V. Barashenkov, E.V. Zemlyanaya, Existence threshold for the ac-driven 
damped nonlinear Schrodinger solitons, arXiv:patt-sol/9906001 v1 28 May 1999.
\bibitem{Glebov-Kiselev-Lazarev}
Glebov S.G., Kiselev O.M.,  Lazarev V.A. Birth of soliton during passage 
through local resonance. 
Proceedings of Steklov Mathematical Institute. Suppl.1, 2003, S84-S90.
\bibitem{SGOK}
Kiselev O.M., Glebov S.G. Scattering of solitons on resonance. ArXiv:math-ph/0403038.
\bibitem{JK}
Jeffrey A. and  Kawahara T. Asymptotic methods in nonlinear wave
theory. Pitman Publishing INC, 1982.
\bibitem{Il'in}
Il'in A.M. Matching of Asymptotic Expansions of Solutions of
Boundary Value Problem, AMS, 1992.
\bibitem{Petv-Pokh}
Petviashvili V.I., Pokhotelov O.A. Uedinennye volny v plazme i atmosfere. M.: Energoatomizdat, 1989. 
\bibitem{DEGM}
Dodd R.K., Eilbeck J.C., Gibbon J.D., Morris H.C., Solitons and nonlinear wave equations. Academic Press, 1984.
\bibitem{Keen-McL} 
Keener J.P., McLaughlin D.W., Solitons
under perturbation. Phys. Rev. A. 1977, v.16, N2. 777-790.

\end{thebibliography}
\end{document}